\documentclass{article}
\usepackage{cite}
\usepackage{graphicx}
\usepackage{dcolumn}

\begin{document}

\date{}
\title{On the spectrum of an oscillator in a magnetic field}
\author{Francisco M. Fern\'{a}ndez \thanks{%
E-mail: fernande@quimica.unlp.edu.ar} \\
INIFTA, Divisi\'on Qu\'imica Te\'orica,\\
Blvd. 113 S/N, Sucursal 4, Casilla de Correo 16,\\
1900 La Plata, Argentina.}
\maketitle

\begin{abstract}
We consider the Hamiltonian for a charged particle in a harmonic potential
in the presence of a magnetic field. The most symmetric case depends on one
parameter, the variation of which leads from a spectrum bounded from below
to an unbounded spectrum. At the transition point the spectrum is bounded
from below but each eigenvalue has infinite multiplicity. The algebraic
method proves to be a remarkable tool for the analysis of this quadratic
Hamiltonian.
\end{abstract}

\section{Introduction}

\label{sec:intro}

The operator $S_{B}=\left( -id/dx_{1}-Bx_{2}/2\right) ^{2}+\left(
-id/dx_{2}+Bx_{1}/2\right) ^{2}$ has received attention because of its role
in superconductivity theory\cite{FH10}. It has a self-adjoint realization
and its spectrum $E_{k}=(2k+1)|B|$, $k=0,1,\ldots $, is not discrete because
each eigenvalue has infinite multiplicity\cite{H13}. A closely related model
has been chosen to study the dynamics of a charged particle in a quadratic
potential in the presence of a uniform magnetic field\cite{RK09,RCR14}. Both
models are particular examples of linear operators that are quadratic
functions of coordinates and momenta.

In a series of papers, it was shown that the algebraic method\cite{FC96} is
one of the simplest and most straightforward ways of studying quadratic
Hamiltonians\cite{F15a,F15b,F16a,F16b,F16c}. This approached is based on the
fact that the regular or adjoint matrix representation of the Hamiltonian
operator in the basis set of coordinates and momenta provides essential
information about the spectrum of the operator. Such matrix representation
appears in the treatment of the dynamics of quadratic Hamiltonians\cite
{RK05, RK09,RCR14,F16b}.

The purpose of this paper is the application of the algebraic method to the
most symmetric form of the Hamiltonian that describes a charged particle in
a quadratic potential in the presence of a uniform magnetic field. By a
continuous deformation of this Hamiltonian through a suitable model
parameter it becomes $S_{B}$ and the spectrum changes dramatically. We show
that the algebraic method is a most useful tool for the description of such
phase transition in the spectrum of the operator.

In section~\ref{sec:algebraic_method} we briefly describe the algebraic
method, in section~\ref{sec:model} we apply it to the chosen model and in
section~\ref{sec:conclusions} we summarize the main results and draw
conclusions.

\section{The algebraic method}

\label{sec:algebraic_method}

In this section we outline the algebraic method discussed in previous papers%
\cite{F15a,F15b,F16a,F16b,F16c}. The model discussed in section~\ref
{sec:model} is a particular case of a quadratic Hamiltonian of the form
\begin{equation}
H=\sum_{i=1}^{2K}\sum_{j=1}^{2K}\gamma _{ij}O_{i}O_{j},
\label{eq:H_quadratic}
\end{equation}
where $\left\{ O_{1},O_{2},\ldots ,O_{2K}\right\} =\left\{
x_{1},x_{2},\ldots ,x_{K},p_{1},p_{2},\ldots ,p_{K}\right\} $, $%
[x_{m},p_{n}]=i\delta _{mn}$, and $[x_{m},x_{n}]=[p_{m},p_{n}]=0$.
Here we restrict ourselves to the case of Hermitian operators
$H^{\dagger }=H$ but the approach applies also to non-Hermitian
ones\cite{F16c}. The algebraic method is particularly useful for
the analysis of the spectrum of $H$ because it satisfies the
commutation relations
\begin{equation}
\lbrack H,O_{i}]=\sum_{j=1}^{2K}H_{ji}O_{j}.  \label{eq:[H,Oi]}
\end{equation}
For this reason it is possible to obtain an operator of the form
\begin{equation}
Z=\sum_{i=1}^{2K}c_{i}O_{i},  \label{eq:Z}
\end{equation}
such that
\begin{equation}
\lbrack H,Z]=\lambda Z.  \label{eq:[H,Z]}
\end{equation}
The operator $Z$ is important for our purposes because if $\left| \psi
\right\rangle $ is an eigenvector of $H$ with eigenvalue $E$ then $Z\left|
\psi \right\rangle $ is eigenvector of $H$ with eigenvalue $E+\lambda $:
\begin{equation}
HZ\left| \psi \right\rangle =(E+\lambda )Z\left| \psi \right\rangle .
\label{eq:HZ|Psi>}
\end{equation}

It follows from equations (\ref{eq:[H,Oi]}), (\ref{eq:Z}) and (\ref{eq:[H,Z]}%
) that
\begin{equation}
(\mathbf{H}-\lambda \mathbf{I})\mathbf{C}=0,  \label{eq:HC=-lambdaC}
\end{equation}
where $\mathbf{H}$ is a $2K\times 2K$ matrix with elements $H_{ij}$, $%
\mathbf{I}$ is the $2K\times 2K$ identity matrix and $\mathbf{C}$ is a $%
2K\times 1$ column matrix with elements $c_{i}$. There are nontrivial
solutions for all values of $\lambda $ that are roots of the characteristic
polynomial $p(\lambda )=\det (\mathbf{H}-\lambda \mathbf{I})$. $\mathbf{H}$
is called the adjoint or regular matrix representation of $H$ in the
operator basis set $\{O_{1},O_{2},\ldots ,O_{2K}\}$\cite{FC96}. This matrix
is closely related to the fundamental matrix that proved to be useful in
determining the conditions under which a PT-symmetric elliptic quadratic
differential operator with real spectrum is similar to a self-adjoint
operator\cite{CGHS12}.

In the case of an Hermitian operator we expect all the roots $\lambda _{i}$,
$i=1,2,\ldots ,2K$ of the characteristic polynomial $p(\lambda )$ to be
real. These roots are obviously the natural frequencies of the
quantum-mechanical system (the actual quantum-mechanical frequencies being
linear combinations of them). It follows from equation (\ref{eq:[H,Z]}) that
\begin{equation}
\lbrack H,Z^{\dagger }]=-\lambda Z^{\dagger },  \label{eq:[H,Z+]}
\end{equation}
where $Z^{\dagger }$ is a linear combination like (\ref{eq:Z}) with
coefficients $c_{i}^{*}$. This equation tells us that if $\lambda $ is a
real root of $p(\lambda )=0$, then $-\lambda $ is also a root. Obviously, $Z$
and $Z^{\dagger }$ are a pair of annihilation-creation or ladder operators
because, in addition to (\ref{eq:HZ|Psi>}), we also have
\begin{equation}
HZ^{\dagger }\left| \psi \right\rangle =(E-\lambda )Z^{\dagger }\left| \psi
\right\rangle .  \label{eq:HZ+|Psi>}
\end{equation}

If $A$ is a quadratic Hermitian operator that satisfies
\begin{eqnarray}
\left[ H,A\right]  &=&0,  \nonumber \\
\left[ A,O_{i}\right]  &=&\sum_{j=1}^{2K}A_{ji}O_{j},  \label{eq:[H,A]=0}
\end{eqnarray}
then it follows from the Jacobi identity
\begin{equation}
\left[ H,\left[ A,O_{i}\right] \right] +\left[ O_{i},\left[ H,A\right]
\right] +\left[ A,\left[ O_{i},H\right] \right] ,
\end{equation}
that
\begin{equation}
\mathbf{HA}-\mathbf{AH}=\mathbf{0},  \label{eq:[mH,mA]=0}
\end{equation}
where $\mathbf{A}$ is the adjoint matrix representation for $A$.

\section{The model}

\label{sec:model}

The starting point is the quadratic Hamiltonian
\begin{equation}
H=\frac{p_{1}^{2}}{2m_{1}}+\frac{p_{2}^{2}}{2m_{2}}+\frac{k_{1}}{2}x_{1}^{2}+%
\frac{k_{2}}{2}x_{2}^{2}+\omega \left( x_{1}p_{2}-x_{2}p_{1}\right) ,
\label{eq:H_gemeral}
\end{equation}
with the customary commutation relations $[x_{i},x_{j}]=0$, $%
[p_{i},p_{j}]=0 $, $[x_{j},p_{k}]=i\hbar \delta _{jk}$. The case $%
m_{1}=m_{2}=m$ was considered to be a simple model for a particle in a
rotating anisotropic harmonic trap or a charged particle in a fixed harmonic
potential in a magnetic field\cite{RK09,RCR14}.

We can reduce the number of free parameters by means of the change of
variables
\begin{equation}
\left( x_{1},x_{2}\right) \rightarrow \left( Lx,Ly\right) ,\;\left(
p_{1},p_{2}\right) \rightarrow \left( \frac{\hbar }{L}p_{x},\frac{\hbar }{L}%
p_{y}\right) ,\;L^{2}=\frac{\hbar }{\sqrt{m_{1}k_{1}}}.
\end{equation}
The resulting dimensionless Hamiltonian
\begin{eqnarray}
\frac{2}{\hbar \omega _{1}}H &=&p_{x}^{2}+\frac{p_{y}^{2}}{\mu }%
+x^{2}+ky^{2}+b\left( xp_{y}-yp_{x}\right) ,  \nonumber \\
\omega _{1} &=&\sqrt{\frac{k_{1}}{m_{1}}},\;\mu =\frac{m_{2}}{m_{1}},\;k=%
\frac{k_{2}}{k_{1}},\;b=\frac{2\omega }{\omega _{1}},
\label{eq:H_general_dimensionless}
\end{eqnarray}
has just three free parameters instead of the original five.

The dynamics of this model was studied in terms of two parameters ($%
k_{x}/\omega ^{2}$ and $k_{y}/\omega ^{2}$) in the case $\mu =1$\cite
{RK09,RCR14}. Here we just consider the most symmetric case $\mu =k=1$ in
order to reduce the number of free parameters to a minimum. Our
dimensionless Hamiltonian will be
\begin{eqnarray}
H &=&H_{0}+bL_{z},  \nonumber \\
H_{0} &=&p_{x}^{2}+p_{y}^{2}+x^{2}+y^{2},  \nonumber \\
L_{z} &=&b\left( xp_{y}-yp_{x}\right) .  \label{eq:H_present}
\end{eqnarray}
This operator is parity-invariant: $PHP=H$, $P\left( x,y,p_{x},p_{y}\right)
\rightarrow \left( -x,-y,-p_{x},-p_{y}\right) $, and the unitary
transformation $U\left( x,y,p_{x},p_{y}\right) \rightarrow \left(
y,x,p_{y},p_{x}\right) $ changes $b$ into $-b$.

The matrix representation of the Hamiltonian (\ref{eq:H_present}) is
\begin{equation}
\mathbf{H}=i\left(
\begin{array}{llll}
0 & -b & 2 & 0 \\
b & 0 & 0 & 2 \\
-2 & 0 & 0 & -b \\
0 & -2 & b & 0
\end{array}
\right)
\end{equation}
The four eigenvalues of $\mathbf{H}$, $\lambda _{1}=-2-b,\lambda
_{2}=2-b,\lambda _{3}=b-2,\lambda _{4}=2+b$, are associated to the operators
\begin{eqnarray}
Z_{1} &=&-y+p_{x}-i(x+p_{y}),  \nonumber \\
Z_{2} &=&y+p_{x}+i(x-p_{y}),  \nonumber \\
Z_{3} &=&y+p_{x}-i(x-p_{y}),  \nonumber \\
Z_{4} &=&-y+p_{x}+i(x+p_{y}),  \label{eq:Z_i(b)}
\end{eqnarray}
respectively. Note that, first, $Z_{1}^{\dagger }=Z_{4}$, $Z_{2}^{\dagger
}=Z_{3}$, second, the pair of ladder operators $\left( Z_{1},Z_{4}\right) $
commutes with the other pair $\left( Z_{2},Z_{3}\right) $ as expected,
third, none of the ladder operators depends on $b$.

The ground-state eigenvalue and eigenfunction are
\begin{eqnarray}
E_{00} &=&2,  \nonumber \\
\psi _{00}(x,y) &=&\frac{1}{\sqrt{\pi }}\exp \left[ -\frac{1}{2}\left(
x^{2}+y^{2}\right) \right] .  \label{eq:psi_00}
\end{eqnarray}
Since $\psi _{00}$ is annihilated by $Z_{1}$ and $Z_{3}$ then $Z_{2}$ and $%
Z_{4}$ are the creation (raising) operators. When $|b|<2$ the spectrum is
bounded from below because $\lambda _{2}>0$ and $\lambda _{4}>0$. On the
other hand, if $|b|>2$ the spectrum is unbounded. In fact, the spectrum is
given by
\begin{equation}
E_{mn}=2+(b+2)m+(2-b)n,\;m,n=0,1,\ldots ,  \label{eq:E_mn(b)}
\end{equation}
and the unnormalized eigenfunctions are
\begin{equation}
\psi _{mn}(x,y)=Z_{2}^{m}Z_{4}^{n}\psi _{00}(x,y).  \label{eq:psi_mn(b)}
\end{equation}

At $b=2$ there is a phase transition between bounded-from-below
and unbounded discrete spectrum. Therefore, the critical point
$b=2$ appears to be interesting. At this point the eigenvalues
$\lambda _{2}$ and $\lambda _{3}$ coalesce $\left( \lambda
_{2}=\lambda _{3}=0\right) $, however the eigenvalues remain real
in all the range of values of the parameter $b$. Other problems
studied earlier exhibit exceptional points at which two real
eigenvalues coalesce and become a pair of complex conjugate
numbers\cite{F15a,F15b,F16a,F16b,F16c}. At an exceptional point
the adjoint matrix becomes defective and some eigenvectors
disappear; that is to say: some ladder operators disappear. The
present case is different as shown below.

When $b=2$ the Hamiltonian operator becomes
\begin{equation}
H=(x+py)^{2}+(y-px)^{2},
\end{equation}
that is exactly the operator $S_{B}$ discussed in the introduction for $B=2$%
. Since $\left[ H,Z_{2}\right] =0$ and $\left[ H,Z_{3}\right] =0$ the
eigenfunctions $\psi _{mn}(x,y)$ are degenerate for all $m=0,1,\ldots $. In
other words, every eigenvalue $E_{mn}=2+4n$ has infinite multiplicity as
stated by Helfer\cite{H13}. Some of the first normalized eigenfunctions are:
\begin{eqnarray}
\psi _{01} &=&\frac{\left( y+ix\right) }{\sqrt{\pi }}\exp \left[ -\frac{1}{2}%
\left( x^{2}+y^{2}\right) \right] ,  \nonumber \\
\psi _{10} &=&\frac{\left( -y+ix\right) }{\sqrt{\pi }}\exp \left[ -\frac{1}{2%
}\left( x^{2}+y^{2}\right) \right] ,  \nonumber \\
\psi _{11} &=&\frac{\left( 1-x^{2}-y^{2}\right) }{\sqrt{\pi }}\exp \left[ -%
\frac{1}{2}\left( x^{2}+y^{2}\right) \right] .  \label{eq:psi_mn_cases}
\end{eqnarray}

The ladder operators $Z_{i}$ are independent of $b$ because $H$, $H_{0}$ and
$L_{z}$ commute. Therefore, their matrix representations $\mathbf{H}$, $%
\mathbf{H}_{0}$ and $\mathbf{L}_{z}$ also commute as shown in general by
equation (\ref{eq:[mH,mA]=0}). For this reason there is a set of
eigenvectors common to the three matrices which is reflected in the
commutation relations
\begin{eqnarray}
\left[ H_{0},Z_{1}\right] &=&-2Z_{1},\;\left[ H_{0},Z_{2}\right]
=2Z_{2},\;\left[ H_{0},Z_{3}\right] =-2Z_{3},\;\left[ H_{0},Z_{4}\right]
=2Z_{4},  \nonumber \\
\left[ L_{z},Z_{1}\right] &=&-Z_{1},\;\left[ L_{z},Z_{2}\right]
=-Z_{2},\;\left[ L_{z},Z_{3}\right] =Z_{3},\;\left[ L_{z},Z_{4}\right]
=Z_{4},  \label{eq:[H0,Zi]/[Lz,Zi]}
\end{eqnarray}
from which it follows that
\begin{eqnarray}
H_{0}\psi _{mn} &=&\left[ 2+2(m+n)\right] \psi _{mn},  \nonumber \\
L_{z}\psi _{mn} &=&(m-n)\psi _{mn}.  \label{eq:H0psimn/Lz_psi_mn}
\end{eqnarray}
These equations clearly explain the results obtained above.

\section{Conclusions}

\label{sec:conclusions}

In this paper we want to draw attention to the symmetric version
of the model for a charged particle in a quadratic potential in
the presence of a uniform magnetic field\cite{RK09,RCR14} because
by suitable deformation it becomes the differential operator
$S_{B}$ discussed by other authors\cite{FH10,H13}. In terms of the
only independent model parameter $b$ the spectrum changes from
being bounded from below ($|b|<2$) to being unbounded ($|b|>2$).
The phase transition takes place at $b=2$ where
$H(b=2)=S_{B}(B=2)$ and the
spectrum is bounded from below but each eigenvalue has infinite multiplicity%
\cite{H13}.

Another purpose of this paper is to show that the algebraic method is a
simple and efficient tool for the analysis of the spectra of quadratic
Hamiltonians. Note that all the results derived in the preceding section
follow from the eigenvalues and eigenvectors of the regular or adjoint
matrix representation of the Hamiltonian operator in the basis set of
coordinates and momenta operators.


\begin{thebibliography}{99}
\bibitem{FH10}  S. Fournais and B. Helffer, \textsl{Spectral methods in
surface superconductivity}, (Springer Science \& Business Media, New York,
2010).

\bibitem{H13}  B. Helffer, \textsl{Spectral theory and its applications},
(Cambridge University Press, Cambridge, 2013).

\bibitem{RK09}  R. Rossignoli and A. M. Kowalski, ''Stability, complex
modes, and nonseparability in rotating quadratic potentials'', Phys. Rev. A
\textbf{79}, 062103 (2009).

\bibitem{RCR14}  L. Rebon, N. Canosa, and R. Rossignoli, ''Dynamics of
entanglement between two harmonic modes in stable and unstable regimes'',
Phys. Rev. A \textbf{89}, 042312 (2014).

\bibitem{FC96}  F. M. Fern\'{a}ndez and E. A. Castro, \textsl{Algebraic
Methods in Quantum Chemistry and Physics}, Mathematical Chemistry Series,
(CRC, Boca Raton, New York, London, Tokyo, 1996).

\bibitem{F15a}  F. M. Fern\'{a}ndez, ''Algebraic Treatment of PT -Symmetric
Coupled Oscillators'', Int. J. Theor. Phys. \textbf{54}, 3871-3876 (2015).

\bibitem{F15b}  F. M. Fern\'{a}ndez, ''Algebraic treatment of a simple model
for the electromagnetic self-force'', arXiv:1509.00002 [quant-ph].

\bibitem{F16a}  F. M. Fern\'{a}ndez, ''Symmetric quadratic Hamiltonians with
pseudo-Hermitian matrix representation'', Ann. Phys. \textbf{369}, 168-176
(2016).

\bibitem{F16b}  F. M. Fernandez, "Algebraic treatment of the Pais-Uhlenbeck
oscillator and its PT-variant'',

\bibitem{F16c}  F. M. Fernandez, "Algebraic treatment of non-Hermitian
quadratic Hamiltonians'', arXiv:1605.01662v3 [quant-ph].

\bibitem{RK05}  R. Rossignoli and A. M. Kowalski, "Complex modes in unstable
quadratic bosonic forms'', Phys. Rev. A \textbf{72}, 032101 (2005).

\bibitem{CGHS12}  E. Calliceti, S. Graffi, M. Hitrik, and J. Sj\"{o}strand,
''Quadratic PT -symmetric operators with real spectrum and similarity to
self-adjoint operators'', J. Phys. A \textbf{45}, 444007 (2012).
\end{thebibliography}
\end{document}